\documentclass[twocolumn,traditabstract]{aa}

\usepackage[nonamebreak]{natbib}
\usepackage[stable]{footmisc}
\usepackage{amsmath}
\usepackage{txfonts}
\usepackage{placeins}
\bibpunct{(}{)}{;}{a}{}{,}
\usepackage{graphicx}
\usepackage{epstopdf}
\usepackage{ifthen}
\usepackage{overpic}
\usepackage{rotating}
\usepackage{epsfig}
\usepackage{multirow}
\usepackage{array}
\usepackage{color}
\usepackage[usernames,dvipsnames]{xcolor}
\usepackage{psfrag}
\usepackage[T1]{fontenc}
\usepackage{url}

\def\ben{\begin{enumerate}}
\def\een{\end{enumerate}}
\def\bi{\begin{itemize}}
\def\ei{\end{itemize}}
\def\be{\begin{equation}}
\def\ee{\end{equation}}
\def\bea{\begin{eqnarray}}
\def\eea{\end{eqnarray}}

\begin{document}

 \title{Probing changes of dust properties along a chain of solar-type prestellar
  and protostellar cores in Taurus with NIKA\thanks{Based on observations carried out under project number 146-13 with the IRAM 30m Telescope. 
IRAM is supported by INSU/CNRS (France), MPG (Germany) and IGN (Spain).}}

\author{
A.~Bracco \inst{\ref{CEA}}
\and P.~Palmeirim \inst{\ref{UniPorto},\ref{LAM}}
\and  Ph.~Andr\'e \inst{\ref{CEA}}
\and  R.~Adam \inst{\ref{LPSC},\ref{OCA}}
\and  P.~Ade \inst{\ref{Cardiff}}
\and  A.~Bacmann \inst{\ref{IPAG}}
\and  A.~Beelen \inst{\ref{IAS}}
\and  A.~Beno\^it \inst{\ref{Neel}}
\and  A.~Bideaud \inst{\ref{Neel}}
\and  N.~Billot \inst{\ref{IRAME}}
\and  O.~Bourrion \inst{\ref{LPSC}}
\and  M.~Calvo \inst{\ref{Neel}}
\and  A.~Catalano \inst{\ref{LPSC}}
\and  G.~Coiffard \inst{\ref{IRAMF}}
\and  B.~Comis \inst{\ref{LPSC}}
\and  A. D'Addabbo \inst{\ref{Neel},\ref{Roma}}
\and  F.-X.~D\'esert \inst{\ref{IPAG}}
\and  P.~Didelon \inst{\ref{CEA}}
\and  S.~Doyle \inst{\ref{Cardiff}}
\and  J.~Goupy \inst{\ref{Neel}}
\and  V.~K{$\ddot{\rm o}$}nyves \inst{\ref{CEA}}
\and  C.~Kramer \inst{\ref{IRAME}}
\and G.~Lagache \inst{\ref{LAM}}
\and  S.~Leclercq \inst{\ref{IRAMF}}
\and  J.F.~Mac\'ias-P\'erez \inst{\ref{LPSC}}
\and  A.~Maury\inst{\ref{CEA}}
\and  P.~Mauskopf \inst{\ref{Cardiff},\ref{Arizona}}
\and  F.~Mayet \inst{\ref{LPSC}}
\and  A.~Monfardini \inst{\ref{Neel}}
\and  F.~Motte \inst{\ref{IPAG}}
\and  F.~Pajot \inst{\ref{IAS}}
\and  E.~Pascale \inst{\ref{Cardiff}}
\and  N.~Peretto \inst{\ref{Cardiff}}
\and  L.~Perotto \inst{\ref{LPSC}}
\and  G.~Pisano \inst{\ref{Cardiff}}
\and  N.~Ponthieu \inst{\ref{IPAG}}
\and  V.~Rev\'eret \inst{\ref{CEA}}
\and  A.~Rigby \inst{\ref{Cardiff}}
\and  A.~Ritacco \inst{\ref{LPSC},\ref{IRAME}}
\and  L.~Rodriguez \inst{\ref{CEA}}
\and  C.~Romero \inst{\ref{IRAMF}}
\and  A.~Roy \inst{\ref{CEA}}
\and  F.~Ruppin \inst{\ref{LPSC}}
\and  K.~Schuster \inst{\ref{IRAMF}}
\and  A.~Sievers \inst{\ref{IRAME}}
\and  S.~Triqueneaux \inst{\ref{Neel}}
\and  C.~Tucker \inst{\ref{Cardiff}}
\and  R.~Zylka \inst{\ref{IRAMF}}}

\institute{
Laboratoire AIM, CEA/IRFU, CNRS/INSU, Universit\'e Paris Diderot, CEA-Saclay, 91191 Gif-Sur-Yvette, France 
  \label{CEA}\\ 
\email{andrea.bracco@cea.fr}
\and
Instituto de Astrof\'isica e Ci{\^e}ncias do Espa\c{c}o, Universidade  
do Porto, CAUP, Rua das Estrelas, PT4150-762 Porto, Portugal
\label{UniPorto}
\and Laboratoire de Physique Subatomique et de Cosmologie, Universit\'e Grenoble Alpes, CNRS/IN2P3, 53, avenue des Martyrs, Grenoble, France
  \label{LPSC}
  \and
  Laboratoire Lagrange, Universit\'e C\^ote d'Azur, Observatoire de la C\^ote d'Azur, CNRS, Blvd de l'Observatoire, CS 34229, 06304 Nice cedex 4, France
  \label{OCA}
  \and
Institut de RadioAstronomie Millim\'etrique (IRAM), Grenoble, France
  \label{IRAMF}
\and
Astronomy Instrumentation Group, University of Cardiff, UK
  \label{Cardiff}
\and
Institut d'Astrophysique Spatiale (IAS), CNRS and Universit\'e Paris Sud, Orsay, France
  \label{IAS}
\and
Institut N\'eel, CNRS and Universit\'e Grenoble Alpes, France
  \label{Neel}
\and
Institut de RadioAstronomie Millim\'etrique (IRAM), Granada, Spain
  \label{IRAME}
\and
Dipartimento di Fisica, Sapienza Universit\`a di Roma, Piazzale Aldo Moro 5, I-00185 Roma, Italy
  \label{Roma}
\and
Univ. Grenoble Alpes, CNRS, IPAG, F-38000 Grenoble, France 
  \label{IPAG}
    \and
Aix Marseille Universit\'e, CNRS, LAM (Laboratoire d'Astrophysique de Marseille) UMR 7326, 13388, Marseille, France
  \label{LAM}
\and
School of Earth and Space Exploration and Department of Physics, Arizona State University, Tempe, AZ 85287
  \label{Arizona}
\and
Universit\'e de Toulouse, UPS-OMP, Institut de Recherche en Astrophysique et Plan\'etologie (IRAP), Toulouse, France
  \label{IRAP}
\and
CNRS, IRAP, 9 Av. colonel Roche, BP 44346, F-31028 Toulouse cedex 4, France 
  \label{IRAP2}
\and
University College London, Department of Physics and Astronomy, Gower Street, London WC1E 6BT, UK
  \label{UCL}
}

\date{Accepted: 23 June 2017}

  \abstract{
 
The characterization of dust properties in the interstellar medium
(ISM) is key for understanding the physics and chemistry of star
formation. Mass estimates are crucial to determine gravitational
collapse conditions for the birth of new stellar objects in molecular
clouds. However, most of these estimates rely on dust models that need
further observational constraints to capture the relevant parameters
variations depending on the local environment: from clouds to
prestellar and protostellar cores. 

We present results of a new study of dust
emissivity changes based on millimeter (mm) continuum data obtained with the
NIKA camera at the IRAM-30m telescope. Observing dust emission at 1.15 mm and 2 mm
allows us to constrain the dust emissivity index, $\beta$, in the
Rayleigh-Jeans tail of the dust spectral energy distribution (SED) far from
its peak emission, where the contribution of other
parameters (i.e. dust temperature) is more important.  
Focusing on the Taurus molecular cloud, one of the most famous
low-mass star-forming regions in the Gould Belt, we analyze the
emission properties of several distinct objects in the B213
filament. This sub-parsec size region is of particular interest
since it is characterized by the presence of a collection of
evolutionary stages of early star formation: three prestellar cores,
two Class-0/I protostellar cores and one Class-II object. We are therefore able to compare dust properties
among a sequence of sources that likely derive from the same parent filament.

By means of the ratio of the two NIKA channel-maps, we show that in
the Rayleigh-Jeans approximation, $\beta_{\rm RJ}$ varies among the objects:
it decreases from prestellar cores ($\beta_{\rm RJ}$$\sim$2) to protostellar
cores ($\beta_{\rm RJ}$$\sim$1) and the Class-II object ($\beta_{\rm
  RJ}$$\sim$0). For one prestellar and two protostellar cores, we produce a robust study using available {\it Herschel} data to
constrain the dust temperature of the sources. By using the Abel
transform inversion technique we get accurate radial temperature
profiles that allow us to obtain radial $\beta$ profiles. We find
systematic spatial variations of $\beta$ in the protostellar cores that is not
observed in the prestellar core. While in the former case $\beta$
decreases toward the center (with $\beta$ varying between 1 and 2), in the latter it
remains constant ($\beta = 2.4 \pm 0.3$). Moreover, the dust emissivity index
appears anticorrelated with the dust temperature. We discuss
the implication of these results in terms of dust grain evolution
between pre- and protostellar cores.  

}


\titlerunning{Changes of dust properties along a chain of solar-type prestellar
  and protostellar cores with NIKA}

\authorrunning{A. Bracco et al.}
\maketitle

\section{Introduction}\label{sec:intro}

The sub-millimeter (sub-mm) dust continuum images obtained with the
{\it Herschel} space observatory unveiled the ubiquitous filamentary structure of
molecular clouds in the interstellar medium (ISM) \citep[e.g.][]{Menshchikov2010,Molinari2010,Hill2011}. Despite the open
debate about the filament formation process in the magnetized and turbulent ISM,
filaments are now conceived as crucial pieces of the puzzle of
low-mass star formation \citep[see][for a review]{Andre2014}.
{\it Herschel} data show that
dense filaments are the preferred loci within molecular clouds where prestellar cores form and eventually
evolve into protostars and pre-main sequence objects \citep[e.g.][]{Andre10,Konyves2015}. 
The progenitors of solar-type stars are believed to go through an evolutionary sequence that
consists of five main stages \citep{Lada1987,
  Andre2000}. When a self-gravitating prestellar core becomes
gravitationally unstable the protostellar stage begins. Depending on
the fraction of accreted mass from the protostellar envelope ($M_{\rm
  env}$) onto the central protostellar object ($M_{\star}$), two protostar stages can be distinguished
(Class-0, $M_{\rm env}>M_{\star}$, and Class-I, $M_{\rm env}<M_{\star}$, \citealp{Andre1993}), before most of the mass finally accumulates 
into the central pre-main sequence star, which also spawns a protoplanetary disk evolving into a debris disk 
(Class-II and Class-III, \citealp{Greene1994}).
At some point along this evolutionary sequence, dust grains are expected to undergo a rapid growth, 
eventually leading to the formation of pebbles and planetesimals. 
Constraining the process of grain growth, which ultimately produces planet formation within protoplanetary 
disks, is the subject of active observational research \citep[e.g.][]{Testi2014}.

In this context, two major issues in low-mass star formation studies are 
1) likely evolution of dust properties from the diffuse ISM to dense cores to protoplanetary disks, 
and 2) accurate mass and column density estimates of star-forming structures.  
Many observational works tackling these issues rely on fitting the
spectral energy distribution (SED) of thermal dust emission at
far-infrared (FIR) and sub-mm wavelengths, where dust emission is usually optically thin
\citep{Hildebrand1983} and therefore a good tracer of the total mass  and column density (of dust and gas)  
integrated along the line of sight. The bulk of the interstellar dust mass is made of big dust grains (BG) in thermal
equilibrium with the interstellar radiation field (ISRF). These grains emit
as greybodies with an equilibrium temperature of few tens of Kelvin, depending on the local environment \citep{Boulanger1996,
  Lagache1998}. Given one BG population,
its emission spectrum (in the optically thin regime) as a function of frequency is
of the form
\begin{equation}\label{eq:sed}
I_{\nu}=\tau_{\nu} B_{\nu}(T_{\rm d})=\kappa_{\nu}\Sigma B_{\nu}(T_{\rm d}),
\end{equation}
where $I_{\nu}$ is the specific brightness (intensity), $\tau_{\nu}$ is
the frequency-dependent optical depth, $B_{\nu}(T_{\rm d})$ is the
Planck function at the temperature $T_{\rm d}$ (the equilibrium temperature of the
grains), $\kappa_{\nu}$ is the dust opacity (in cm$^2$g$^{-1}$), and $\Sigma$ is the gas surface density distribution. The latter can be written as $\Sigma=\mu_{{\rm H}_2}m_{\rm H}N_{{\rm
    H}_2}$, given the mean molecular
weight per hydrogen molecule $\mu_{{\rm H}_2}$, the mass
of the hydrogen atom $m_{\rm H}$, and the column density of molecular
hydrogen $N_{{\rm H}_2}$. The dependence of $\kappa_{\nu}$ on the
frequency is usually expressed as
\begin{equation}\label{eq:emi}
\kappa_{\nu}=\kappa_0\left(\frac{\nu}{\nu_0}\right)^{\beta},
\end{equation}
where $\kappa_{0}$ is the emissivity cross-section per gram of dust
and gas at the frequency
$\nu_0$, and $\beta$ is the dust emissivity index.  The shape of the
observed SED mainly depends on $T_{\rm d}$ and $\beta$, while its
offset level on $\Sigma$. The dust temperature
controls the frequency position of the SED peak. It is set by the strength of the
ISRF and by the emitting grain size distribution, since
bigger grains have a lower equilibrium temperature than smaller grains
\citep{Mathis1977,Weingartner2001}. Less obvious is the dependence of
$\beta$ on dust properties \citep{Draine2006}. It relates to the dust structure and
composition, which determine both the optical/UV absorption cross-section and the emission cross-section of the grains, and the
frequency-dependent efficiency to emit radiation (see Eq.~(\ref{eq:emi})) \citep{Compiegne2011,Jones2017}.
It follows that the mass estimate of a {\it dusty}
source depends on all possible assumptions related to these parameters. 
\begin{figure*}
\begin{tabular}{c c}
\includegraphics[width=9.cm]{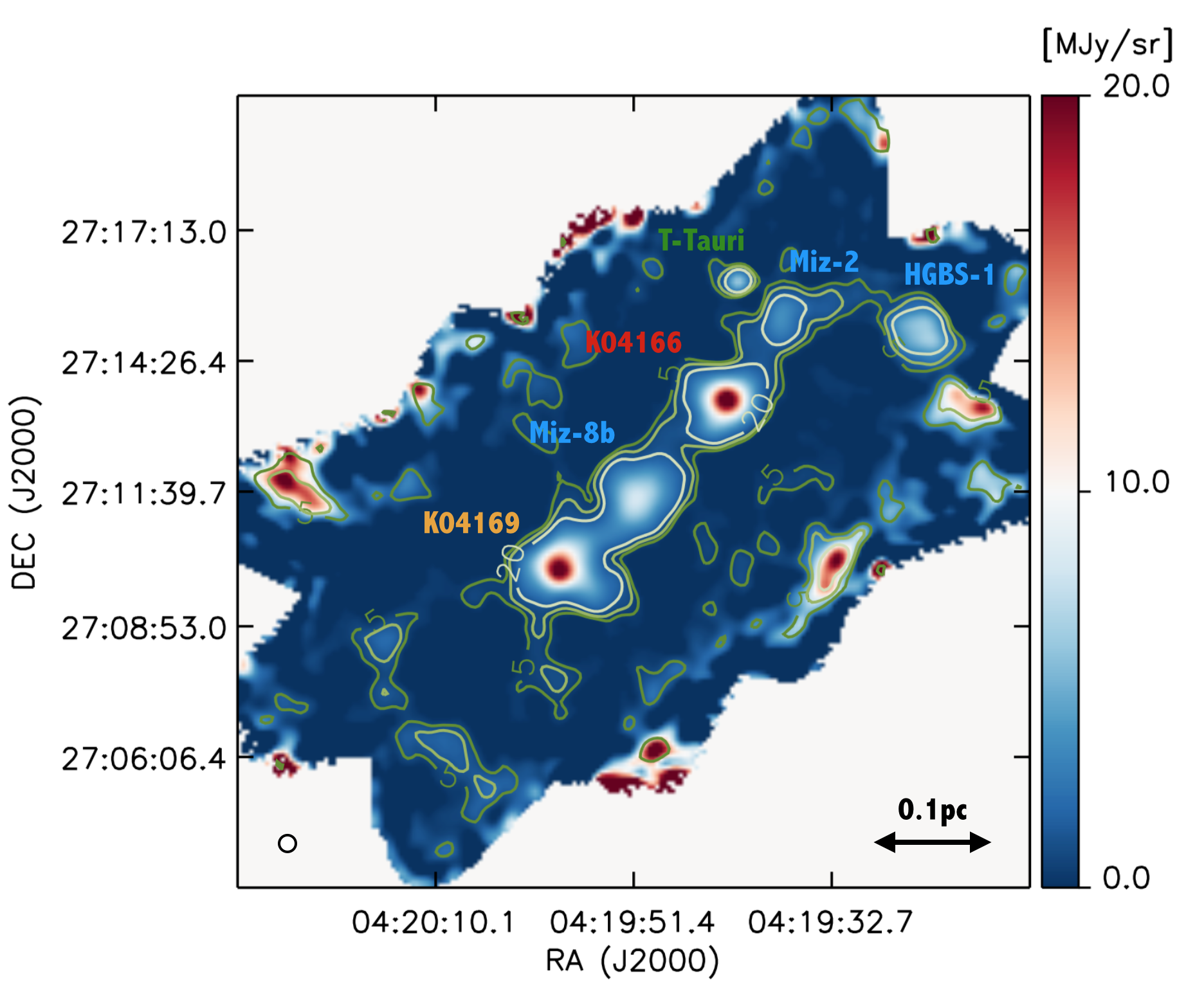}
 & \includegraphics[width=9.cm]{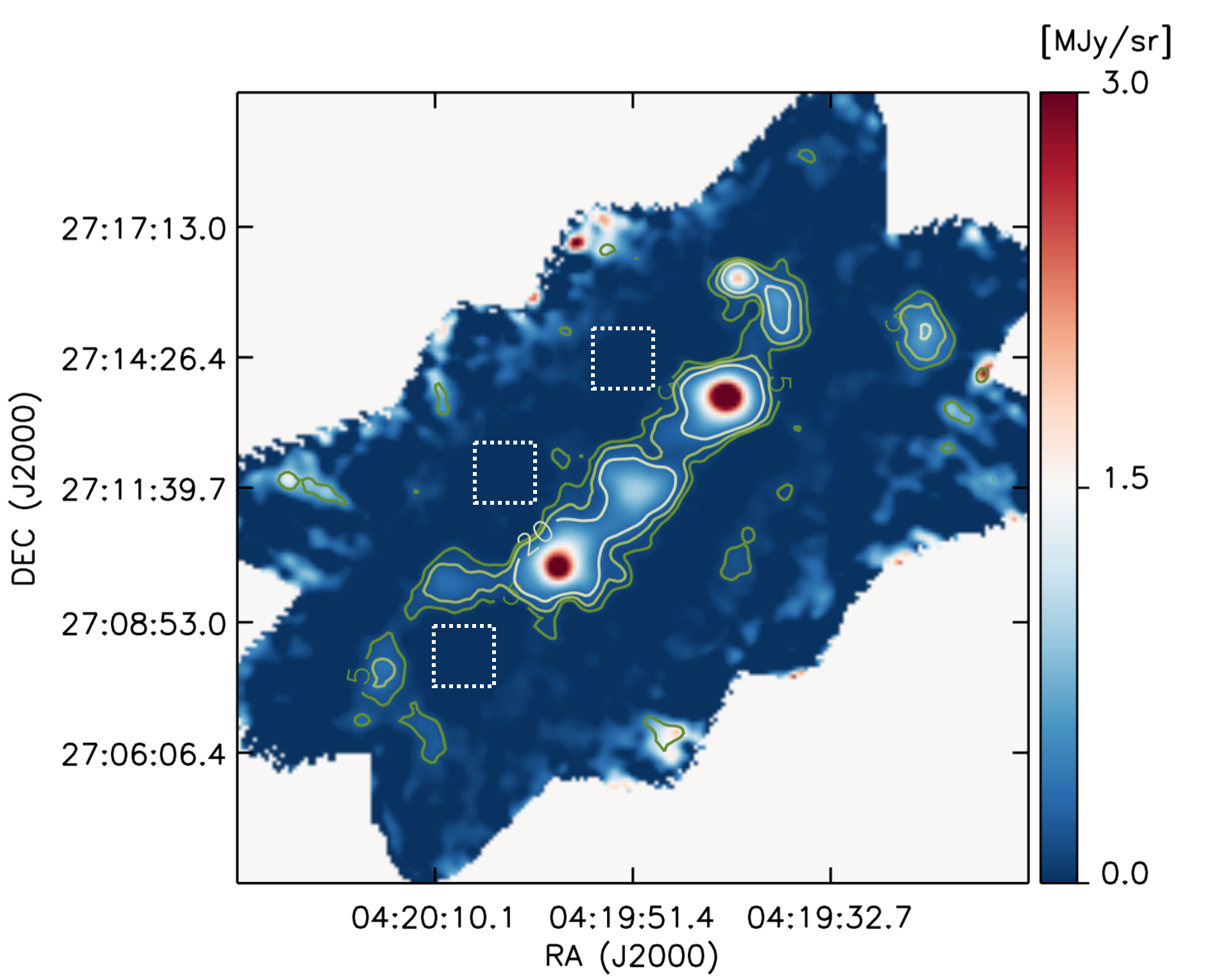}\\
\end{tabular}
   \caption[]{Dust continuum maps of a portion of the B213 filament obtained with the NIKA
        camera on the IRAM 30~m telescope at 1.15 mm (left panel) and 2 mm (right panel).
        The two maps are shown after smoothing to a common resolution
        of $24''$ (see the circle on the left-bottom part of the left
        panel). The green contours represent the signal-to-noise levels
        (5, 10, 20) at each wavelength. The labels in the left panel
        correspond to the sources analyzed in this work and described in
        Table~\ref{tab:sources}. The three squares on the right
      delimit the regions where we estimate the rms intensity of the
      two maps at $24''$, which corresponds to $0.75$ MJy/sr and $0.09$ MJy/sr at 1.15 mm and 2 mm
      respectively.}
      \label{fig:Nika}
\end{figure*}

\begin{table*}
\hspace{2.5cm}
\begin{tabular}{| c | c | c | c | c |} 
  \hline \hline	
    &  &  & &\\ 		
 Object & Evolutionary & RA & DEC & Abbreviation in \\ 
 & Stage &(J2000)& (J2000) & text and figures  \\ \hline 
 &  &  & &\\ 
HGBS-J041937.7+271526$^{\rm a,e}$ &  Prestellar & 04h 19m 37.7s
         &+27$^{\circ}$15\arcmin\,20.0\arcsec &  Miz-2\\
HGBS-J041923.9+271453$^{\rm e}$ &  Prestellar & 04h 19m 23.9s
         &+27$^{\circ}$14\arcmin\,53.0\arcsec & HGBS-1 \\
Miz-8b$^{\rm a,c}$ &  Prestellar & 04h 19m 51.0s
         &+27$^{\circ}$11\arcmin\,42.2\arcsec & Miz-8b \\
K04166$^{\rm b}$& Class-0/I & 04h 19m 42.9s
         &+27$^{\circ}$13\arcmin\,38.8\arcsec & K04166\\
K04169$^{\rm b}$& Class-0/I &  04h 19m 58.9s
         &+27$^{\circ}$10\arcmin\,00.5\arcsec & K04169 \\ 
J04194148+2716070$^{\rm d}$ & T-Tauri & 04h 19m 41.5s
         &+27$^{\circ}$16\arcmin\,07.0\arcsec & T-Tauri \\ 
 &  &  & &\\ 
  \hline \hline 
\end{tabular}
\caption{Description of the analyzed prestellar and protostellar objects in the B213
  filament \citep[a,b,c,d,e
  -][respectively]{Mizuno1994,Motte2001,Tatematsu2004,Davis2010,Marsh2016}. Celestial
  coordinates are given according to the corresponding positions in the NIKA maps.}
\label{tab:sources}
\end{table*}
Several works at FIR and sub-mm wavelengths studied how properties of dust emission vary, both at large scales in the more diffuse ISM \citep{Bracco2011,planck2013-p06b,Ysard2015} and at small
scales in star-forming regions \citep{Stepnik2003,Paradis2010,Ysard2013,Sadavoy2013,Roy2014}. 
However, disentangling the contributions of $\beta$ and
$T_{\rm d}$ variations in dust SEDs is a difficult task in the FIR and
sub-mm domains. One of the keys to this problem is to probe dust emission at longer
wavelengths in the Rayleigh-Jeans portion of the SED,
where temperature effects are less important. Thus, the millimeter (mm) range
is crucial to investigate dust emissivity index variations.   
While some models still assume $\beta$ to be constant (equal to 2 for crystalline dust
grains, \citet{Draine2007,Compiegne2011}), recent
observational results between $1$ and $3$ mm
\citep[i.e.,][]{Schnee2014,Sadavoy2016} revealed $\beta$
variations in the Orion molecular cloud, which may be caused by possible dust evolution in
protostars. Changes of $\beta$ at smaller scales in
protoplanetary disks were also observed and interpreted as consequence of dust
grain-growth in disk formation \citep{Draine2006,Guilloteau2011}.

In this paper we present new results from millimeter continuum observations of a portion 
of the Taurus molecular cloud, obtained in winter 2014 with the Institut de Radio Astronomie
Millimetrique (IRAM) 30-m telescope using the dual-band capability of the New IRAM KID Arrays \citep[NIKA,][]{Monfardini2011,Calvo2013,Adam2014,Catalano2014}.
This prototype instrument enabled us to simultaneously map at $\sim$1.15 mm
(band-1) and $\sim$2 (band-2) mm a 8' x 5' segment of the B213
filament in the Taurus molecular cloud, where three prestellar cores,
two Class-0/I protostellar cores and one Class-II object are embedded. We propose an original
analysis of dust-emissivity-index spatial variations associated with several
evolutionary stages of early star formation for objects that likely
originated from the same parent filament.
 
In this work we make use of both NIKA and {\it Herschel}
data, which we combine avoiding biases due to the different data acquisition process of the two
instruments. We base our analysis on the ratio map between the two
NIKA channels. Putting constraints on the dust temperature of the
observed sources by means of the {\it Herschel} data only, we are able to 
infer radial changes of $\beta$ that correlate with the presence of protostars. 

This study highlights the capabilities of the NIKA camera, which will
be the starting point for a new observational campaign in star forming
regions with the advent of the NIKA2 instrument in both intensity and polarization \citep{Calvo2016}. 

The paper is organized as follows. Section~\ref{sec:B213}
describes the B213 filament in the Taurus molecular cloud and the main
sources we focus on. In Sect.~\ref{sec:nika} we present NIKA and the
data reduction process. Section~\ref{sec:ratio} introduces the ratio
map between the two NIKA bands, which guides our data analysis. The
study of the dust-emissivity variations among the various targets
inferred from the ratio map is presented in Sect.~\ref{sec:betavar},
while the discussion of the results is in
Sect.~\ref{sec:discussion}. We conclude this paper with summary and
prospects in Sect.~\ref{sec:conclusion}.  
\begin{figure}
     \includegraphics[width=9.cm]{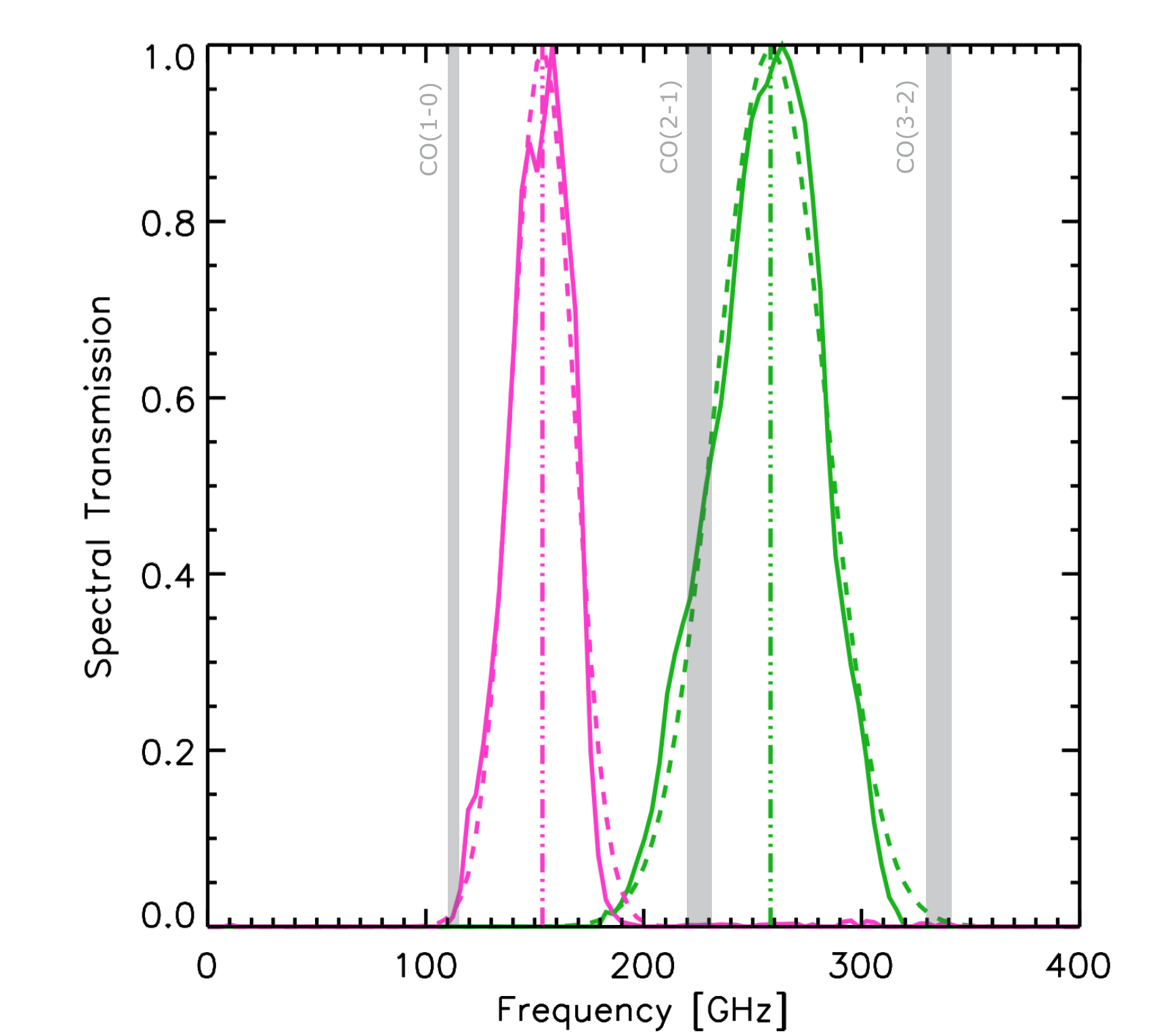} 
      \caption[]{Spectral transmission of the two NIKA bands. The
        dashed curves represent Gaussian fits to the bandpasses. The
        central peaks, marked with vertical-dotted lines, are at $(260 \pm
        25)$ GHz ($\sim$1.15 mm) in green and $(150 \pm 14)$ GHz  ($\sim$2 mm) in pink. The grey shades show the CO transitions
        that fall into the NIKA bandpasses.}
      \label{fig:bandpasses}
\end{figure}
\section{B213 filament and its embedded objects}\label{sec:B213}
The Taurus molecular cloud is one of the closest ($d~\sim~140$~pc, \citet{Elias1978})
and well studied low-mass star-forming region in the Gould Belt.
Most of its young stars are grouped in elongated patterns that are closely associated with a network of prominent parallel
filaments \citep[e.g.][]{Schneider1979,Hartmann2002}. B213 is one of the most dominant filaments 
that is actively forming stars \citep{Schmalzl2010,Li2012}, where
several embedded prestellar and protostellar cores are clearly detected along its crest \citep{Onishi2002,Marsh2016}.

{\it Herschel} Gould Belt Survey (HGBS) observations revealed the
density and dust temperature structure of B213 with unprecedented detail.
\citet{Palmeirim2013} argued that the filament is self-gravitating and currently contracting quasi-statically toward its longest axis while accreting material from the surrounding environment.
Its radial density profile approaches power-law
behavior, $\rho \propto r^{-2}$ at large radii, and the temperature profile significantly drops toward the crest. 
They report central average values for the dust temperature and gas
volume density in the filament of 11.8~K and $\sim 5 \times 10^4$~cm$^{-3}$, respectively.

In the NIKA maps of the segment of B213 displayed in Fig.~\ref{fig:Nika} we can see six bright sources, which
are listed in Table~\ref{tab:sources}. Three of them are prestellar
cores (hereafter, Miz-8b, Miz-2, HGBS-1 \citep[see][]{Mizuno1994,Marsh2016} in blue in Fig.~\ref{fig:Nika}). Miz-8b is also detected in
${\rm N}_{2}{\rm H}$$^+(1-0)$ observations with the Nobeyama 45 m telescope
\citep{Tatematsu2004}. The other three sources are not prestellar
cores but bright point-like objects
detected by {\it Herschel} PACS at 70$\mu$m. Two of them, those
neighbouring Miz-8b, are protostellar cores classified as
Class-0/I objects (hereafter, K04166 and K04169\footnote{ The first letter “K” of
  the adopted names recalls IRAS sources selected by \citet{Kenyon1990,Kenyon1993}.} \citep{Motte2001,Tafalla2010}), which are
known to drive bipolar outflows \citep{Bontemps1996,Tafalla2004,SantiagoGarcia2009}. As shown in
Fig.~5 of \citet{Motte2001} by the envelope-mass versus
bolometric-luminosity plot, the two sources are border-line Class-0/Class-I objects with strong near-infrared emission
\citep{Kenyon1993, Furlan2008}.

In the north-western direction of the field of view a more
evolved object (J04194148+2716070) classified as a Class-II T-Tauri star
can be also seen \citep{Davis2010}.

In-depth analysis of molecular line observations of this field
have revealed the presence of velocity coherent structures, referred to as
``fibers'' \citep{Hacar2013}, which overlap on the plane of the sky towards B213, with typical lengths of $\sim$0.5~pc.
However, we note that the portion of B213 observed with NIKA
corresponds to a single fiber. Thus, the objects discussed in this
paper likely belong to the same parent physical structure.
The fact that these objects are different evolutionary
stages of early star formation evolved from the same structure, makes them an interesting 
case to study how the dust emissivity index varies as star formation proceeds.

\section{Observations and data reduction}\label{sec:nika}

We made use of the dual-band capability of NIKA, which
was used to observe part of B213 in the Taurus molecular
cloud. The transmission curves of the two bands are displayed in
Fig.~\ref{fig:bandpasses}. The central frequencies are $260 \pm 25$
GHz (or $\sim$1.15 mm) for band-1 (with $\sim$190 valid pixels) and $150 \pm 14$ GHz (or
$\sim$2 mm) for band-2 (with $\sim$125 valid pixels). The two NIKA brightness maps, $I_{1\rm mm}$ and $I_{2 \rm mm}$, are shown in Fig.~\ref{fig:Nika} in units of MJy/sr.
Observations were performed during the first NIKA open pool, in February 2014. The observing conditions were stable with a mean zenith
opacity of 0.13 in band-1 and 0.09 in band-2. Uranus was used as primary calibrator.

Given the broad coverage of the NIKA bandpasses (see
Fig.~\ref{fig:bandpasses}), we applied average color
corrections of $0.97$ and $0.91$ to the data in band-1 and band-2,
respectively. These corrections were estimated integrating over the
NIKA bandpasses greybodies for dust emission with
spectral indices, $\beta$, varying between $1$ and $3$ and including the IRAM-30m atmospheric models for the mean opacities referred to above. 

The overall calibration uncertainty, including the
calibrator model uncertainty, was estimated to be 7\%  in band-2 and 12\%
in band-1 \citep{adam2015}. Pointing corrections were checked using
nearby quasars every hour, leading to a pointing rms accuracy better
than $3\arcsec$. The telescope focus was checked about every 3
hours. The effective beam FWHM was measured to be $18.2\arcsec$ and
$12.0\arcsec$ in band-2 and band-1, respectively. However, since in
this work we also make use of {\it Herschel} data between $160\, \mu$m and $350\, \mu$m of B213 (see Sect.~\ref{ssec:betaH}),
the NIKA maps were smoothed to the resolution of $24\arcsec$
  (with $5\arcsec$ pixels) of SPIRE-$350\, \mu$m, assuming Gaussian beams for both NIKA
channels. The error due to the Gaussian assumption of the
  smoothing kernel was found to be negligible compared to considering the
  actual NIKA point spread function of the two bands, and did not
  affect the results of the paper. The rms noise level at $24\arcsec$
  resolution was estimated on the
three source-free square regions in the right panel of
Fig.~\ref{fig:Nika}. It corresponds to $0.75$ MJy/sr in band-1 
and $0.09$ MJy/sr in band-2 \footnote{At the original NIKA resolution
  the rms noise level is $4$ mJy/beam and $1$ mJy/beam at 1.15 mm
  (12$\arcsec$ beam) and at 2 mm (18.2$\arcsec$ beam), respectively.}. 

The signal was mapped using scans made of a set of subscans, separated
by $20\arcsec$ from each other, oriented perpendicular, and oriented
with an angle of $\pm$ $45^\circ$ with respect to the main filament
axis. The scans cover a quasi homogeneous region of
8'$\times$5'. Such scanning strategy allows us to limit residual
stripping in the reduced maps, thanks to the variety in the scanning
angles, and to limit filtering effects by allowing to define the
surface brightness zero level in the external regions of the
filament. The data in the two bands were reduced simultaneously and independently as described in \citet{adam2015}. In brief, the correlated noise (atmospheric and electronic) was removed by averaging and subtracting the detector timelines across the arrays. A small fraction of the scans were flagged due to bad weather conditions and some others were lost because of missing data streams with the telescope position. The overall effective time spent on target was 16.1 hours.

The data processing filtered out the astrophysical signal
on scales larger than the NIKA field of view ($\sim$2 arcmin). The
filtering was the same at both wavelengths \citep{Adam2017}, but the
effective zero level brightness in each map was unconstrained. It was
defined during the mapmaking procedure, in an iterative manner, by
setting to zero the mean of the detector timelines in the low surface brightness
parts of the filament, i.e., where the signal-to-noise ratio in the
maps at their nominal resolutions was
lower than 3. In order to take into account the uncertainty
introduced by the mapmaking procedure, we tested the robustness of our
results against four versions of the maps produced with
different filtering criteria depending on the size of the mask used to
define the background signal. 

\section{Ratio map between the two NIKA bands}\label{sec:ratio}
\begin{figure}
     \includegraphics[width=9.cm]{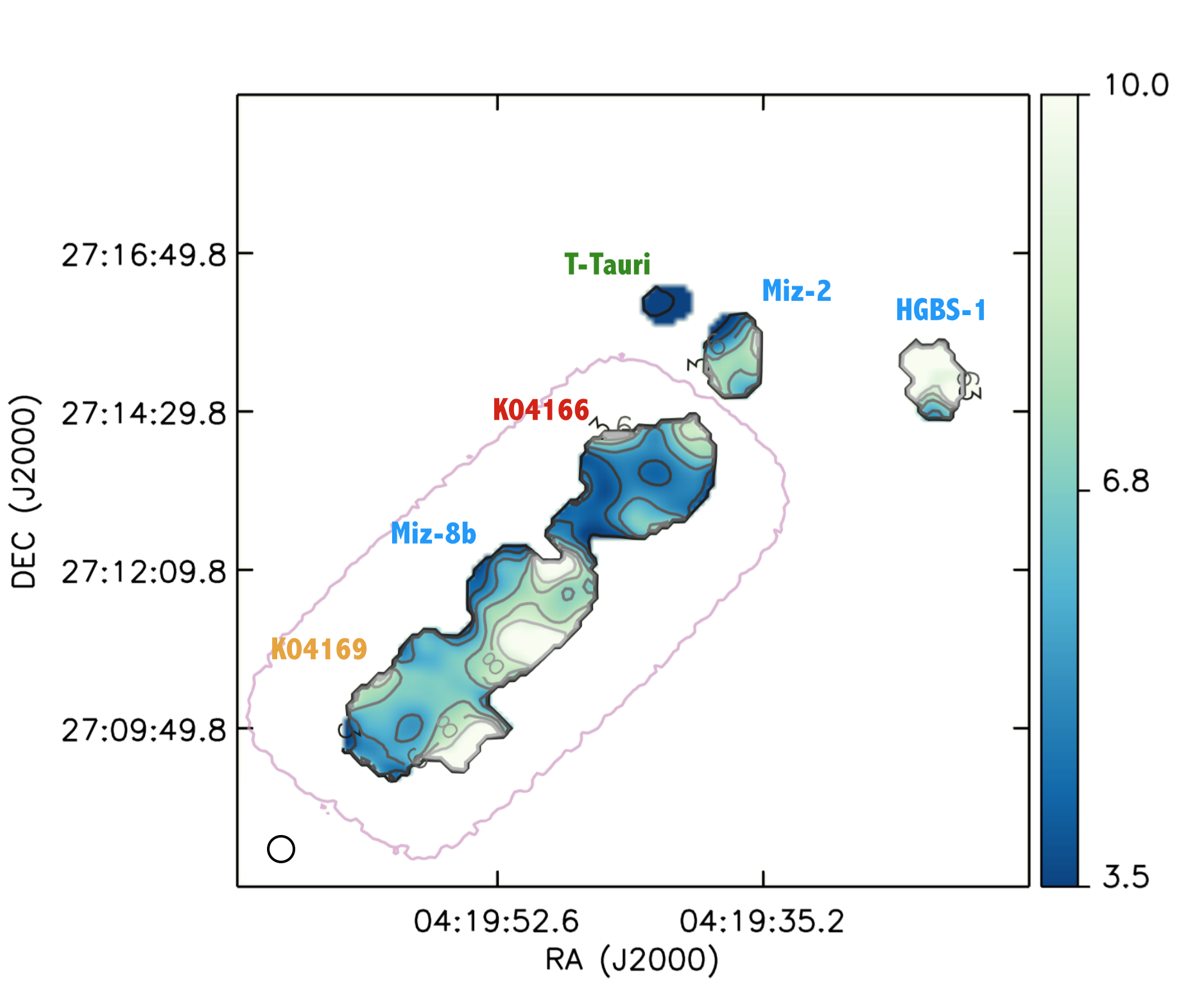} 
      \caption[]{Ratio map between the two NIKA bands, $R_{1,2}=\frac{I_{1\rm mm}}{I_{2\rm
          mm}}$. The unmasked region corresponds to a signal-to-noise
        ratio on the fluxes above $10$ at both wavelengths. Black contours represent
        $R_{1,2}$ equal to 3, 4.5, 5.6, 7, 8, and 9. The purple
        contour corresponds to the part of the field with homogeneous
        observing time $>$ 240 s per pixel (see Fig.~\ref{fig:timemap}). The open
        circle at the bottom left represents the beam size ($24\arcsec$).}
      \label{fig:Ratiomap}
\end{figure}
\begin{figure*}
\begin{tabular}{c c}
\includegraphics[width=9.cm]{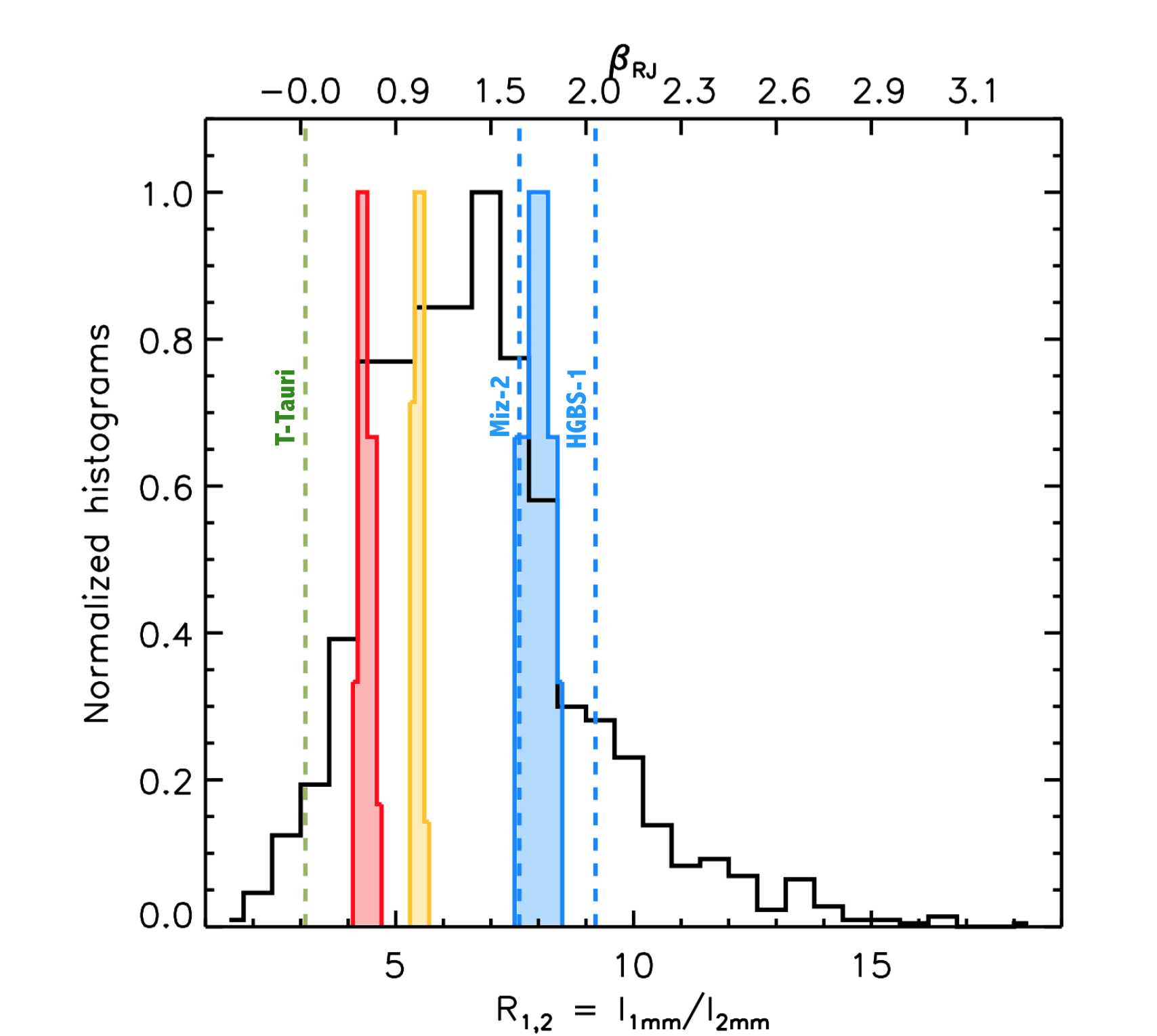}
 & \includegraphics[width=9.cm]{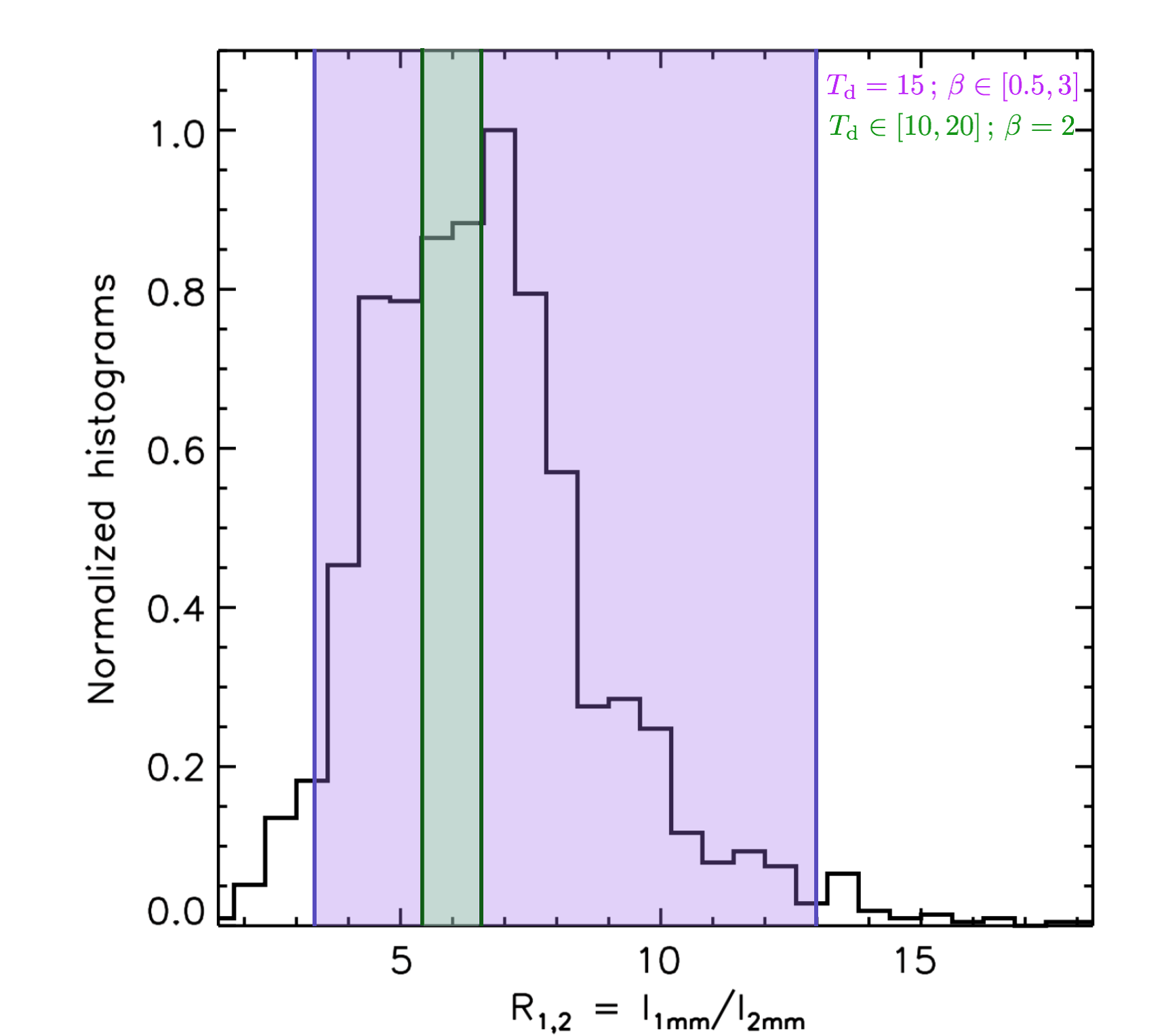}\\
\end{tabular}
      \caption[]{{\it Left}: normalized histogram of $R_{1,2}$ in
        black for all the unmasked pixels in the maps. The coloured normalized histograms represent
        the $R_{1,2}$ distributions within one-beam area ($24\arcsec$) around the
        central coordinates (see Table~\ref{tab:sources}) of K04166
        (red),  K04169 (yellow), and Miz-8b (blue). The vertical
        dashed lines refer to the value of the ratio at the
        peak-intensity position of T-Tauri (green), Miz-2, and HGBS-1 (blue).
          The upper x-axis shows the corresponding $\beta$ values in the
        Rayleigh-Jeans limit (see Sect.~\ref{ssec:betarj}). {\it
          Right}: same as in the left panel with shades that now represent ranges of $R_{1,2}$ corresponding to
         greybodies with 1) a constant temperature $T_{\rm d}$ of
        $15$ K but a varying $\beta$ between $0.5$ and $3$ in purple; 2) a
        constant $\beta$ of $2$ but varying $T_{\rm d}$ between
        $10$ K and $20$ K in green.} 
    \label{fig:RatiosHisto}
\end{figure*}

To put constraints on the dust
emissivity index variations in the mm-domain we study the ratio
between the NIKA surface brightnesses. In Fig.~\ref{fig:Ratiomap} we show the ratio
map, $R_{1,2}=\frac{I_{1\rm mm}}{I_{2\rm mm}}$, where we
masked all pixels with a signal-to-noise ratio lower than 10 in at least one of
the two bands at the resolution of $24\arcsec$. In the image it is possible to appreciate the difference in the value of $R_{1,2}$
between the various objects. In particular, prestellar cores clearly
exhibit a larger value of $R_{1,2}$
compared to the other sources. The lowest value of 
$R_{1,2}$ corresponds to the position of the Class-II T-Tauri object, the most evolved object
in the sample. In order to strengthen the change in $R_{1,2}$ with
evolutionary stage, the left panel of
Fig.~\ref{fig:RatiosHisto} presents four normalized histograms of
$R_{1,2}$, where only the unmasked pixels in the
maps are considered. The histogram corresponding to all unmasked pixels is
shown in black. It peaks at $\sim$7 with a broad distribution skewed toward lower values. In colours we show the normalized histograms of
$R_{1,2}$ within one-beam area ($24\arcsec$) centered around K04166 (red),  K04169 (yellow),
and Miz-8b (blue). For the remaining three objects we only display, with
dashed-vertical lines, the value of $R_{1,2}$ at the peak intensity
positions of the T-Tauri star (green) and the prestellar cores Miz-2 and HGBS-1 (blue). As detailed in
Appendix~\ref{sec:app}, the observing time was not completely homogeneous across the NIKA maps (see
Fig.~\ref{fig:timemap}). For this reason our study mainly
focuses on the sources with the longest observing time, corresponding to K04166,  K04169,
and Miz-8b (see the purple contour in Fig.~\ref{fig:Ratiomap}). Nevertheless Fig.~\ref{fig:RatiosHisto} also shows the
peak $R_{1,2}$ values measured for the other objects. The difference among 
all sources is striking. While prestellar cores present values of
$R_{1,2}$ of $\sim$8, the ratio progressively decreases to $\sim$5.5
and $\sim$4.3 for the protostellar cores and to $\sim$3 for
the Class-II T-Tauri object.

These variations of $R_{1,2}$ in the wavelength regime probed by NIKA are difficult to explain without
invoking changes in the dust emissivity index of the sources. In the right
panel of Fig.~\ref{fig:RatiosHisto} the result of a simple exercise to
test this assessment is shown. Assuming a greybody spectrum for dust
emission (see Eq.~\ref{eq:sed}), we verify the effect on $R_{1,2}$ of significant variations of $T_{\rm d}$ but negligible changes in
$\beta$, and vice-versa. Thus, on top of the normalized histogram of
$R_{1,2}$, we display with green shades the range of values that we
would get with $10 {\rm K} <T_{\rm d}<20 {\rm K}$ for a fixed $\beta=2$, and with purple shades the
range of $R_{1,2}$ for a fixed $T_{\rm d}=15$ K  and $0.5<\beta<3$. The
specific values of the parameters are not critical for
this test. The figure clearly shows the impact of varying
$\beta$ in this wavelength range compared to changes in $T_{\rm
  d}$. The range of $R_{1,2}$ expected from plausible $\beta$ variations allows us
to recover almost $95\%$ of the values observed with NIKA in Fig.~\ref{fig:Ratiomap}. 
    
\section{Variations of the dust emissivity index from the ratio
  map}\label{sec:betavar}

In the following section we derive the dust emissivity index, $\beta$, for
K04166, K04169, and Miz-8b from the ratio map, $R_{1,2}$, introduced
in Sect.~\ref{sec:ratio}. We first
estimate $\beta$ in the Rayleigh-Jeans limit and then
produce a more accurate evaluation using constraints on the dust
temperature of the sources based only on {\it Herschel} data.   

\subsection{$\beta$ values in the Rayleigh-Jeans limit}\label{ssec:betarj}

In the previous section we showed how variations of $\beta$ can affect dust emission at long wavelengths
compared to changes in $T_{\rm d}$. This suggests that we can, to first order, consider the Rayleigh-Jeans limit of
the Planck function ($h\nu \ll k_{\rm B}T_{\rm d}$, where $h$ and
$k_{\rm B}$ are the Planck's and the Boltzmann's constants,
respectively\footnote{ At 260 GHz the Rayleigh-Jeans limit corresponds to $T_{\rm
    d} \gg 12 {\rm K}$}) to directly deduce the values of $\beta$ from
$R_{1,2}$. Under this assumption we write $B_{\nu}(T_{\rm
  d})\approx 2k_{\rm B}T_{\rm d}\nu^2c^{-2}$, so that, from
Eq.~\ref{eq:sed} and Eq.~\ref{eq:emi}, $\beta$ as a
function of $R_{1,2}$ is
\begin{equation}\label{eq:betaRJ}
\beta_{\rm RJ}=\frac{\log{R_{1,2}}}{\log{(\nu_{1 \rm mm}/\nu_{2 \rm mm})}}-2,
\end{equation} 
where $\nu_{1 \rm mm}$ and $\nu_{2 \rm mm}$ are $\sim$260 GHz and $\sim$150 GHz
respectively and the subscript ``RJ'' stands for Rayleigh-Jeans. 

The top x-axis of the left panel of Fig.~\ref{fig:RatiosHisto} shows
an indicative scale with the corresponding values of $\beta_{\rm
  RJ}$. The overall distribution peaks
at $\beta_{\rm RJ}=1.4$ and ranges from very low values below
$\beta_{\rm
  RJ}$$\sim$0.5 to large values above $\beta_{\rm
  RJ}$$\sim$3. As for $R_{1,2}$, we find a clear difference between the
$\beta_{\rm RJ}$ values for protostellar cores (0.6, 1.1) and for prestellar cores ($\sim$2). Clearly, protostellar cores tend to
have lower values of $\beta_{\rm RJ}$ compared to prestellar cores.

We notice that the lowest values of $\beta_{\rm RJ}$$\sim$0 mostly
relate to the Class-II object, which may be enough optically thick
to justify these low values. However a more detailed and accurate discussion is not
possible since the Class-II object is located in the part of the maps
with varying sensitivity and observing time (see Appendix~\ref{sec:app}). 

\subsection{Radial profiles: from {\it Herschel} dust temperatures to $\beta$}\label{ssec:betaH}

In order to go beyond the simple Rayleigh-Jeans approximation we produce a more detailed analysis, which relies on putting
constraints on the dust temperature structure of each source using
HGBS data for B213 \citep{Palmeirim2013,Marsh2016}. For this detailed
analysis, we only consider K04166, K04169, and Miz-8b, which are the
three objects observed with the most homogeneous sensitivity with NIKA
(cf. Fig.~\ref{fig:timemap} and purple contour in Fig.~\ref{fig:Ratiomap}).

The {\it Herschel} data allow us to estimate the
dust temperature of each source since the PACS and SPIRE
maps probe the peak of the dust SED between $160\,\mu$m
and $500\,\mu$m. 

Our framework is the following: we use {\it Herschel} data at
  wavelengths close to the peak of the dust SED to best evaluate
  $T_{\rm d}$; we make use of NIKA data at mm-wavelengths in the
Rayleigh-Jeans portion of the SED to estimate $\beta$.

An accurate estimate of the temperature for compact sources,
such as those presented in this study, requires some caution. A
single-temperature greybody fit of the dust SED (see Eq.~\ref{eq:sed}) 
provides us with a value of the temperature of a given source averaged along the line of
sight. This approach neglects possible temperature gradients
within the observed objects caused by internal (protostellar case) or
external (prestellar case) radiation fields. Because of this effect,
the mass estimate of the sources may be also affected \citep[e.g.][]{Malinen2011,Roy2014}. 

Thus, to estimate the dust temperature profiles of our three
sources we make use of the reconstruction method described in
\citet{Roy2014}, which consists in an inversion technique based on the
Abel integral transform applied to the {\it Herschel} data between
PACS-$160\,\mu$m and SPIRE-$350\,\mu$m. 

Under the assumption of spherical symmetry for a source embedded in a
uniform radiation field and with radial density profile $\rho(r)$,
the specific intensity $I_{\nu}(p)$ observed at a distance $p$ from
the center of the source can be expressed as
\begin{equation} \label{eq:abelint}
I_{\nu}(p) \cong 2\int^{+\infty}_{p} \rho(r)B_\nu(T_{\rm
  d}(r))\kappa_{\nu}\frac{r{\rm d} r}{\sqrt{r^2-p^2}}.
\end{equation}  
The Abel transform allows us to infer the integrand of the above
equation given the intensity at each frequency as
\begin{equation} \label{eq:abelinv}
\rho(r)B_\nu(T_{\rm
  d}(r))\kappa_{\nu}= -\frac{1}{\pi} \int^{+\infty}_{r} \frac{{\rm
    d}I_{\nu}}{{\rm d} p}\frac{{\rm d} p}{\sqrt{p^2-r^2}}.
\end{equation}  
Assuming spherical symmetry, the derivatives of the radial intensity
profiles are sufficient to estimate $\rho(r)$ and $T_{\rm d}(r)$. Given
$\kappa_{\nu}$ we can simultaneously obtain the
density (or column density, $N_{{\rm H}_2}$, if the line-of-sight integration is considered) and temperature radial profiles by fitting a single
temperature greybody to the {\it
  Herschel} SED between $160\,\mu$m and $350\,\mu$m. We ignore here SPIRE-$500\,\mu$m since we want to get
advantage of the highest possible resolution of the NIKA dual-band data. In this
case the greybody fit of the {\it Herschel} data is performed with
$\beta=2$ and, from \citet{Roy2014}, 
\begin{equation}\label{eq:decs}
\kappa_{\nu}=0.1\times (\lambda/300\,\mu{\rm m})^{-2} \,
{\rm cm}^2/{\rm g}.
\end{equation}
 Nevertheless, we produce temperature radial profiles
considering both $\beta=2$ and $\beta=1.55$, with no significant impact on
the results. We notice that thanks to the SED frequency coverage provided by the HGBS data no degeneracy exists between density
and temperature through the Abel reconstruction. While the shape of
the SED is constrained by the dust temperature, the strength of the sub-mm
continuum emission is set by the density.  
\begin{figure*}
\begin{tabular}{c c}
\includegraphics[width=9.cm]{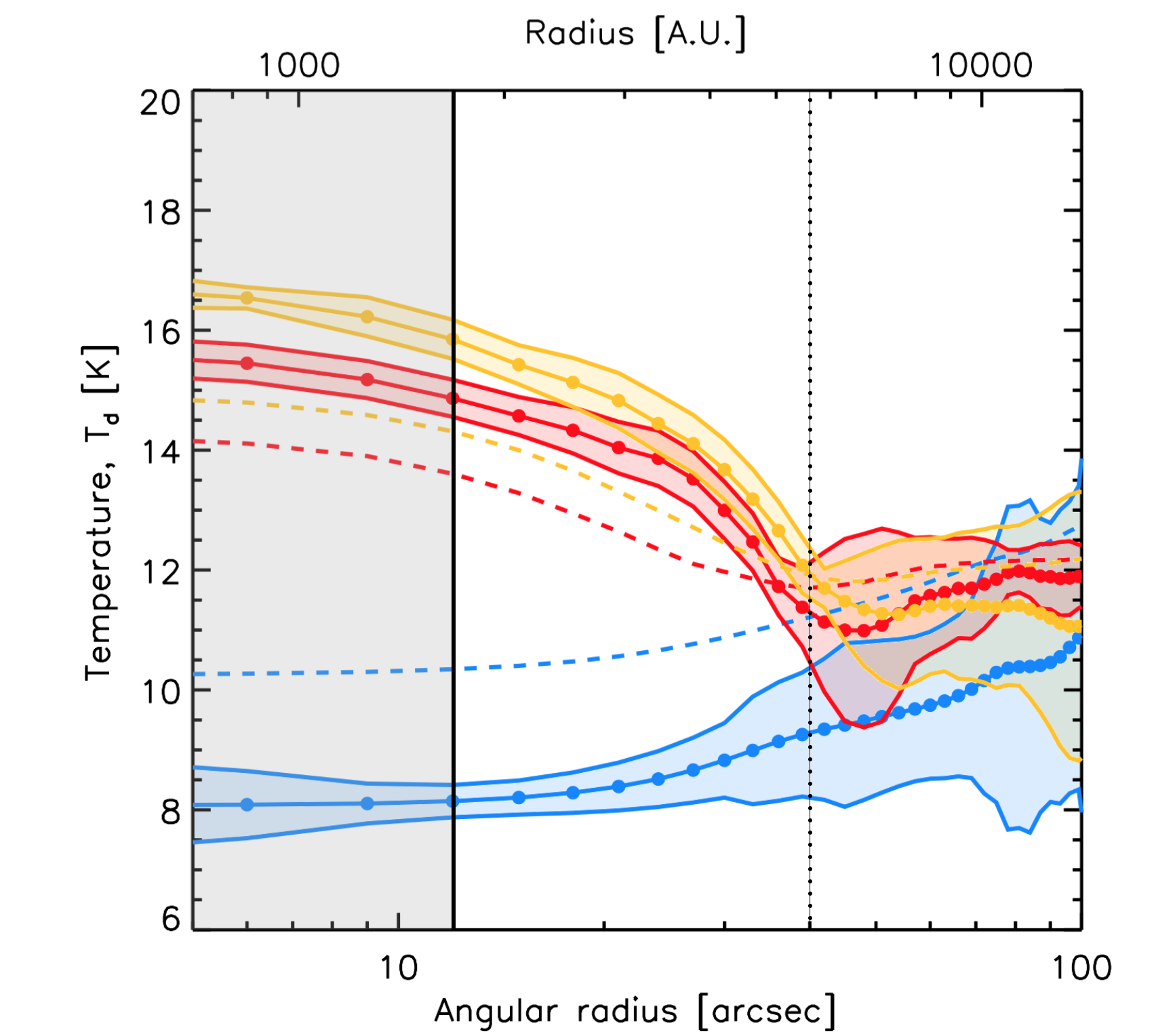}
 & \includegraphics[width=9.cm]{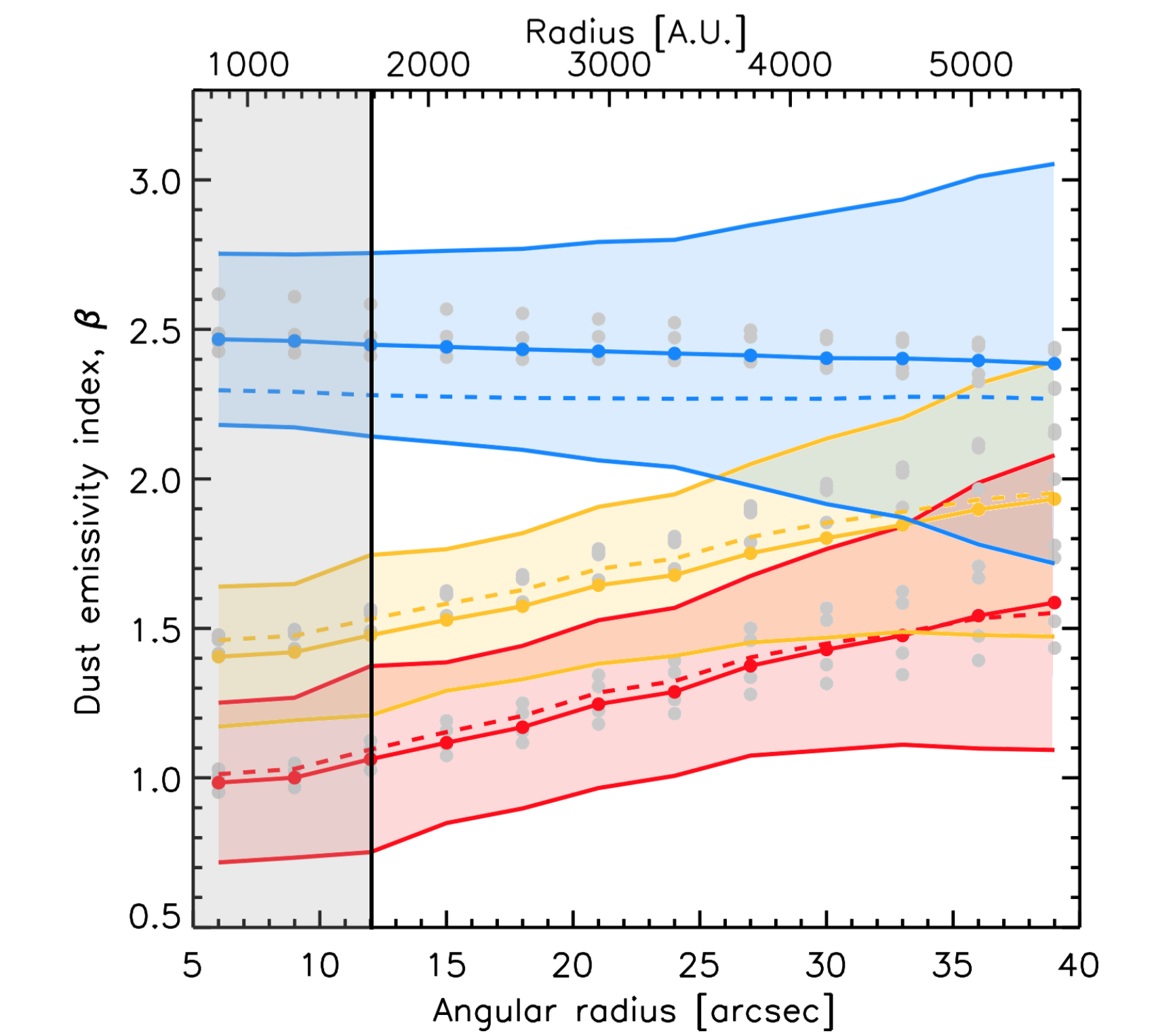}\\
\end{tabular}
 \caption[]{{\it Left}: Dust temperature radial profiles of K04166
        (red), K04169 (yellow), and Miz-8b
       (blue). The dust 
        temperatures were derived from {\it Herschel} Gould Belt
        survey data (http://gouldbelt-herschel.cea.fr/archives/). The
        dashed-coloured lines correspond to line-of-sight averaged temperatures obtained from direct SED fitting, 
        while the dots and the solid lines correspond
        to Abel-reconstructed dust temperatures. The coloured shades represent the 1-$\sigma$ errors of the
        Abel-reconstructed temperatures. {\it Right}: radial profiles of $\beta$ using the
        Abel-reconstructed temperatures. The
        colour-code is the same as in the left panel. The grey dots show the $\beta$ profiles corresponding to four different versions of the maps used
        to quantify systematic effects due to spatial filtering in the
        data. The grey-shaded area in both panels indicates radial
        distances below the half-beam width of $12\arcsec$. The Abel
        transform is only performed for radial distances smaller than
        the dotted-black vertical line in the left panel. On the upper
        x-axis the radial distance is in astronomical units
        assuming a distance of $140$pc for the Taurus molecular cloud.}
       \label{fig:tempbeta}
\end{figure*}
In the left panel of Fig.~\ref{fig:tempbeta} we show the radial profiles of $T_{\rm d}$
with the usual color code corresponding to the three sources under
study. The data points are the Abel-reconstructed temperatures while,  for
comparison, the dashed lines show the line-of-sight SED dust temperatures. As
expected, the main difference between the two cases is at the center
of the sources, where the line-of-sight temperature underestimates the
true value of $T_{\rm d}$ for the protostellar objects and overstimates it for the
prestellar core. On the other hand, the Abel-reconstructed and the line-of-sight
dust temperatures tend to agree at large radii from source center.   

Given these radial temperature profiles, we can 
now go beyond the Rayleigh-Jeans approximation presented in
Sect.~\ref{ssec:betarj} and use $T_{\rm d}(r)$ to perform a more
rigorous estimate of $\beta$ at radius, $r$, from the center of each source, as follows  
\begin{equation}\label{eq:beta}
\beta(r)=\frac{\log{\{R_{1,2}\times B_{\nu_{2 \rm mm}}[T_{\rm d}(r)]/B_{\nu_{1 \rm mm}}[T_{\rm d}(r)]\}}}{\log{(\nu_{1 \rm mm}/\nu_{2 \rm mm})}}.
\end{equation} 
In the right panel of Fig.~\ref{fig:tempbeta} we display the radial profiles of $\beta$ with the same colour code as in the left
panel. The
error bars represent the combination of the 1 $\sigma$ uncertainties derived from
propagating the statistical errors in\footnote{The standard deviation in $R_{1,2}$ is obtained by propagating the errors in the mean
  values of $I_{1 \rm mm}$  and  $I_{2 \rm mm}$ within bins of the
  radial distance.} $R_{1,2}$ and $T_{\rm d}$ and
the errors in the zero-level offsets of the two maps introduced in
Sect.~\ref{sec:nika}. These errors are assumed, in a conservative way, to
be equal to the corresponding rms values (0.75 MJy/sr and 0.09 MJy/sr
at 1.15 mm and 2 mm, respectively). The $\beta$ radial profiles
are limited to an angular radius of $40\arcsec$ in order to only use data with a signal-to-noise
ratio above 10 in both NIKA bands (see Sect.~\ref{sec:ratio}).

In this diagram the different trends between the protostellar cores
and the prestellar core are remarkable and listed in Table~\ref{tab:linfit}.
The prestellar core has a flat profile while both protostellar cores reveal an increasing
trend of $\beta$ with radius, with $\beta$ values that become
consistent with those of the prestellar core at the farthest
distance from the center. This effect is observed both when $T_{\rm
  d}(r)$ is obtained from the Abel inversion technique (coloured data points)
and when it is deduced from the line-of-sight dust temperature (dashed lines).
The mapmaking procedure applied
to the NIKA data may affect the ratio map and the 
value of $\beta$. We thus repeated our analysis with the four different versions of
the maps (see Sect.~\ref{sec:nika}) and we show the corresponding $\beta$ radial profiles with
grey data points in the right panel of Fig.~\ref{fig:tempbeta}. Systematic
effects related to the mapmaking do not significantly change our
results. The different trends observed in the $\beta$ profiles of the three sources are indeed
preserved, although the total uncertainty in the absolute values
increases. The $\beta$ profile of the prestellar core is consistent
with a constant value of $\beta=2.4 \pm 0.3$.

\begin{table}
\center
\begin{tabular}{| c | c | c | } 
  \hline \hline	
    &  &   \\ 		
 Object  & Slope (m) [arcsec$^{-1}$] & Inner $\beta$ value ($\beta_{\rm in}$) \\
  &  &   \\  \hline
  &  &   \\ 
Miz-8b & $0.00 \pm 0.01$ & $ 2.4 \pm 0.3$ \\
K04166& $0.019 \pm 0.009$ & $0.8 \pm 0.2$ \\
K04169& $0.018\pm 0.008$ & $1.3 \pm 0.2$ \\ 
  &  &   \\ 
  \hline \hline 
\end{tabular}
\caption{Parameters of the linear fit, $\beta(r) = m\times r + \beta_{\rm
    in}$, to the $\beta$ profiles shown in the right
  panel of Fig.~\ref{fig:tempbeta}, where $r$ is the angular radius.}
\label{tab:linfit}
\end{table}

\section{Discussion}\label{sec:discussion}
In this section we discuss potential astrophysical biases affecting our
results and provide plausible explanations of the $\beta$
radial profiles shown in Fig.~\ref{fig:tempbeta}. 

First of all, we notice that other types of radio emission, such as
free-free or (gyro-)synchrotron radiation, are negligible compared to dust
emission in this work. The Taurus molecular
cloud is a low-mass star-forming region, differing from those studied in
similar works such as Orion in \citet{Schnee2014}  and in
\citet{Sadavoy2016}, with very few internal ionizing sources.
Furthermore, all of the young stellar objects (YSOs) present in the field of Fig.~\ref{fig:Nika} are low-luminosity
objects ($L_{\rm bol} < 1\, L_{\sun}$). Strong (gyro-)synchrotron emission is quite rare 
among such YSOs and restricted to Class-III objects
(cf. \citealp{Andre1996}). The base of the jets/outflows emanating from Class-0 (or -I) protostars produce weak 
free-free emission from shock-ionized gas in the centimeter range \citep{Anglada1996}, 
but the expected contribution is negligible at mm wavelengths. The total free-free
radio continuum flux scales with the outflow momentum rate and can be estimated 
from Eq. (1) of \citet{Anglada1996} at 5 GHz. Given the outflow momentum rates measured 
by \citet{Bontemps1996} for the protostars K04166 and K04169, the expected free-free 
flux density is $\sim$0.2 mJy at 5 GHz, which scales to an upper limit of about $\sim$0.6 mJy 
at 150 GHz using a power-law frequency dependence with a spectral index of 0.3 for extrapolation (see Table 1 of \citealp{Anglada1996}, for L1551-IRS5 in Taurus).
Considering the NIKA rms noise level of ∼1 mJy/18.2"-beam at 150 GHz, we conclude 
that free-free radio contamination can definitely be neglected in our study, even 
for the protostellar cores. 
Shocked material associated with the outflows from the two
protostars, K04166 and K04169 \citep{Bontemps1996,
  Tafalla2010}, may also have an unusually high dust emissivity
index ($\beta$$\sim$3) \citep{Gueth2003}. However, we do not see evidence
of such $\beta$ values here.

We also made sure that CO contamination to the mm continuum is
not significant for the three sources. In Fig.~\ref{fig:bandpasses} we
show that the only emitting transition that may be relevant for our study is the CO(2-1), which
falls within NIKA band-1. However, using available FCRAO 12CO data
\citep{Narayanan2008}, and integrating the line over the NIKA bandpass, we
estimated that the total CO contamination is less than $1\%$ of the
NIKA surface brightness in the case of the two protostellar cores and even lower in the case
of the prestellar core Miz-8b.   
\begin{figure}
     \includegraphics[width=9.cm]{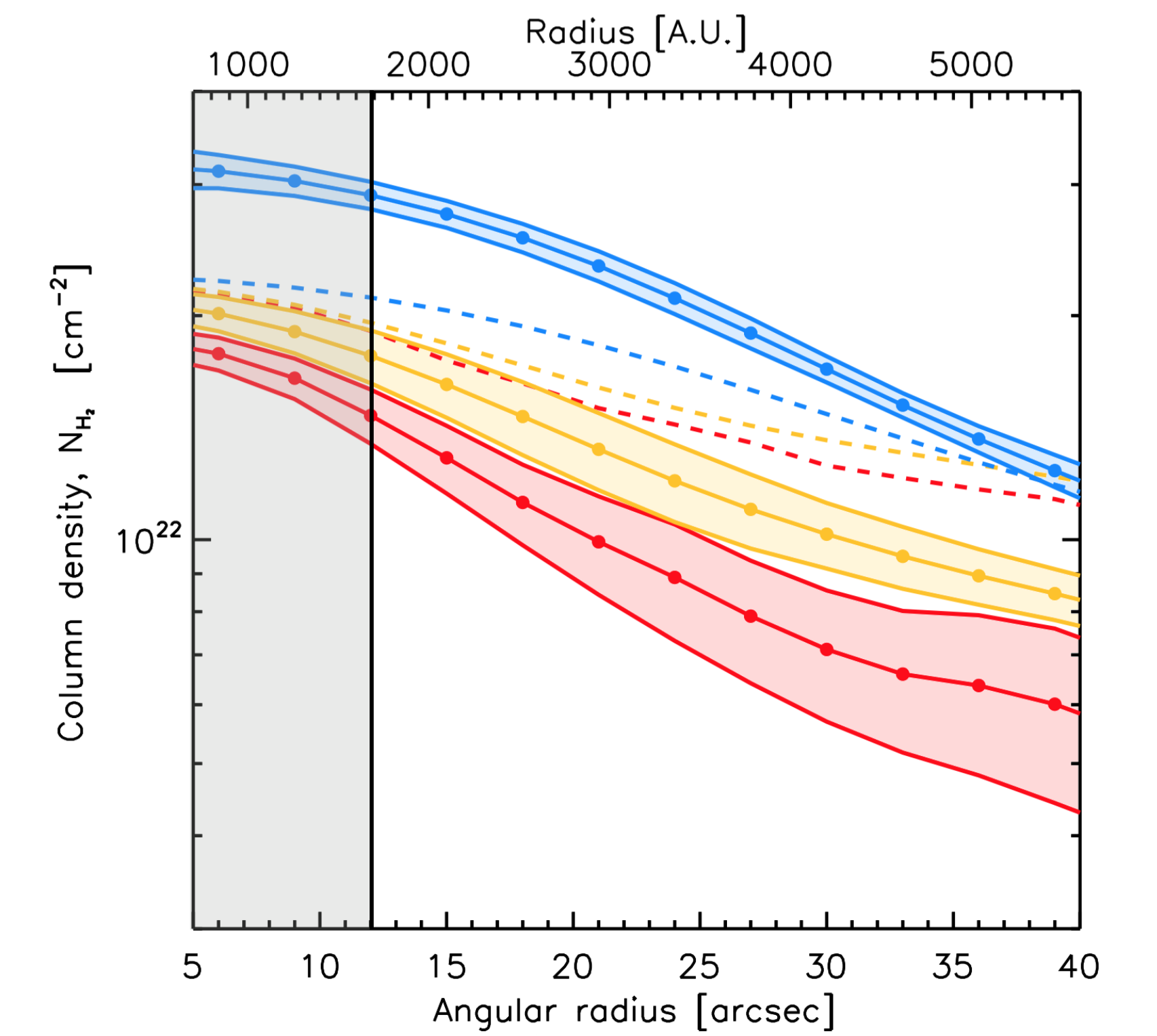} 
      \caption[]{Radial $N_{H_2}$ column density profiles for K04166
        (red), K04169 (yellow), and Miz-8b
       (blue), as derived from {\it Herschel} Gould Belt survey data.       
       The colour-code is the same as in Fig.~\ref{fig:tempbeta} as
       well as the grey-shaded area.}
      \label{fig:nh2}
\end{figure}

We also verified the hypothesis
of optically thin dust emission in both band-1 and band-2.
Given the Abel-reconstructed $N_{{\rm H}_2}$ radial profiles (see Fig.~\ref{fig:nh2}), obtained from the {\it Herschel} data by 
normalizing the greybody fit described in
Sect.~\ref{ssec:betaH}, and given Eq.~\ref{eq:decs}, we can provide a rough
estimate of the optical depth at 1.15 mm for the three objects as follows
\begin{equation}
\tau_{1{\rm mm}}=\kappa_{1{\rm mm}}\mu_{{\rm H}_2}m_{\rm H}N_{{\rm H}_2},
\end{equation}
where $\mu_{{\rm H}_2}=2.8$ \citep{Kauffmann2008}. Using the radial profiles in Fig.~\ref{fig:nh2} the
largest value of $\tau_{1{\rm mm}}$ on a $24\arcsec$-scale obtained for the
prestellar core Miz-8b is $0.001$, which is much smaller than unity. If instead of using $\beta=2$ in
Eq.~\ref{eq:decs} we vary $\beta$ between 1 and 3, results do not change.  

In spite of the approximations made above, these estimates suggest that the
trends seen in the $\beta$ radial profiles of the three sources must be
real and associated to changes in dust properties. 

\subsection{Astrophysical interpretation}\label{ssec:interp}
A number of works already showed 
that dust-grain growth,
up to mm-size, may cause lower values of dust emissivity 
indices (as low as 0) especially toward Class-II objects with protoplanetary disks
\citep[e.g.][]{Guilloteau2011}. Younger Class~0
objects in the Orion molecular cloud were also found with low $\beta$
values ($\sim$0.9) compared to those of the diffuse ISM \citep{Schnee2014},
although a recent work by \citet{Sadavoy2016} revisited this result finding larger values
between 1.7 and 1.8, reporting a lack of evidence for grain
growth in OMC2/3. In this context our results represent an important
contribution to the study of dust evolution at early stages of star
formation. We may potentially be witnessing a change
in the size distribution of dust grains going from the outer parts of protostellar cores, where $\beta$ is compatible with that of the
prestellar core Miz-8b, to the inner part, where the increase in
dust-grain size and change in chemical composition (i.e. aliphatic-rich to
aromatic-rich amorphous hydrocarbons, \citet{Jones2013}) may explain a
larger emissivity of the grains, coupled with the drop of the $\beta$ radial
profiles toward the center. In addition to changes in $\beta$, we
  notice that intrinsic changes in $\kappa_{0}$ (see Eq.~\ref{eq:emi}), caused
by grain growth, may also exist \citep{Kramer2003}. However, this effect cannot be constrained
with NIKA data only, since the ratio $R_{1,2}$ is independent of $\kappa_{0}$ variations.

\begin{figure}
     \includegraphics[width=9.cm]{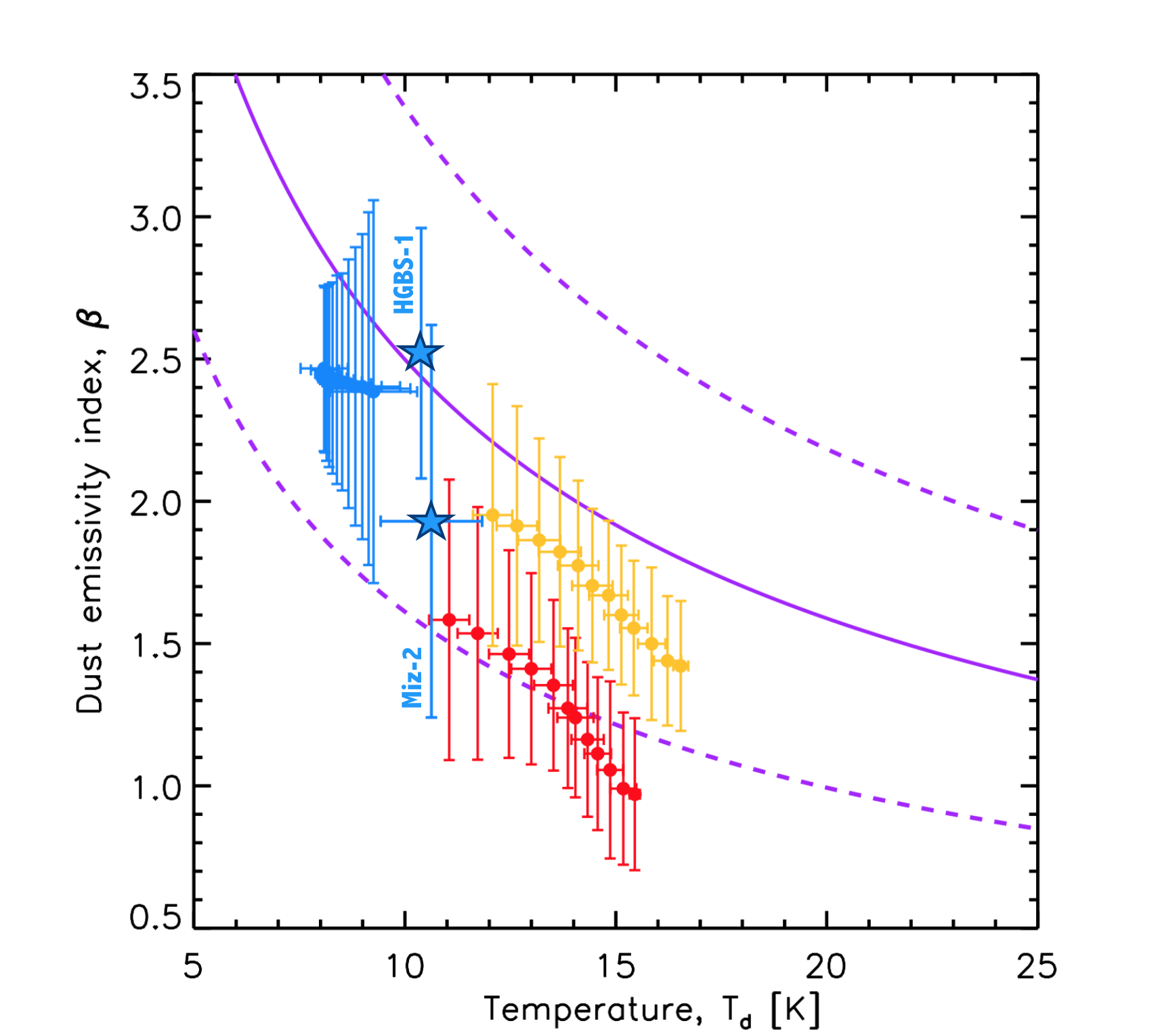} 
      \caption[]{Diagram of $\beta$ against $T_{\rm d}$ for K04166
        (red), K04169 (yellow), and Miz-8b
       (blue). For indication, the peak values for Miz-2 and
       HGBS-1 are also displayed with blue stars. The purple lines represent the empirical law
       (see Sect.~\ref{sec:discussion}) found in
     \citet{Desert2008} (solid line) and the corresponding 1-$\sigma$
     uncertainty (dashed lines).}
      \label{fig:BT}
\end{figure}

Another interpretation might explain our
results. Both from theoretical models and laboratory experiments, it
has been known for a while that because of the microphysics of dust
grains the dust emissivity index in the mm range can depend on
temperature: the higher the dust temperature the lower the
$\beta$ \citep{Agladze1996,Mennella1998, Boudet2005, Meny2007,Coupeaud2011}. In Fig.~\ref{fig:BT} we show the $\beta$-$T_{\rm d}$ diagram relative to
the three sources under study with the Abel-reconstructed dust temperature
profiles. We also display the peak $\beta$ values as a function of the Abel-reconstructued $T_{\rm d}$ for the prestellar cores Miz-2
and HGBS-1 (blue stars in Fig.~\ref{fig:BT}). Overall, a clear anticorrelation between the two parameters is
observed. Previous observational works performing dust SED fits with PRONAOS, ARCHEOPS, {\it Herschel}, and {\it
  Planck} data reported such an anticorrelation across different physical scales and astrophysical
environments \citep[e.g.][]{Dupac2003,Desert2008,Bracco2011,planck2013-p06b}. While
in most of these studies a possible statistical bias due to low signal-to-noise
regions in the maps may mimic the intrinsic anticorrelation between
$\beta$ and $T_{\rm d}$ \citep{Shetty2009}, in the present work the observed trend
is not sensitive to this effect. Indeed, we separately estimated the
two parameters, $\beta$ and $T_{\rm d}$, based on NIKA and {\it
  Herschel} data respectively, without introducing any statistical correlation caused
by noise. 
Nevertheless, as discussed in
\citet{Coupeaud2011}, because of many assumptions in the modelling (i.e., main focus on the disorder of the amorphous state of
silicates), and of various spurious effects in observations (i.e.,
line-of-sight integration, noise), a one-to-one comparison between our results and
theoretical expectations is complicated. Most laboratory experiments find an
anticorrelation between $\beta$ and $T_{\rm d}$ over a wide range in
temperature ($\sim$100K, \citet{Coupeaud2011}). \citet{Agladze1996}
show the dependence between the two parameters in a range more
similar to ours, with values of $\beta$ close to 2.5 at $10$K for {\it
  forsterite} and {\it enstatite} silicate grains and a shallower decrease of $\beta$ with $T_{\rm
  d}$ at higher temperatures compared to our finding. Interestingly, the
results displayed in Fig.~\ref{fig:BT} are in agreement (within $1 \sigma$) with the
empirical power law of the form $\beta(T_{\rm d})= (11.5\pm 3.8) T_{\rm d}^{(-0.660\pm0.054)}$ (purple in the figure), found by
\citet{Desert2008} studying dust properties toward cold clumps in the Galactic
plane with ARCHEOPS data.  

We conclude that intrinsic changes in dust emissivity related to
temperature variations may also play a role in the $\beta$
radial profiles shown in Fig.~\ref{fig:tempbeta}, although a definite
interpretation based on state-of-the-art dust models and laboratory data is yet to be
achieved. We believe that the most likely scenario causing the decrease of $\beta$ toward
the center of protostellar cores is a combination of both
grain growth and dust temperature effects.

\section{Summary \& prospects}\label{sec:conclusion}

Making use of the dual-band capability of NIKA to
simultaneously observe the sky at 1.15 mm and 2 mm, in this paper we 
presented new results on the analysis of dust
emissivity changes in the
B213 filament of the Taurus molecular cloud (at a distance of
$\sim$140pc) . 
The observed region is a $8'\times5'$ portion of a self-gravitating filament in the
process of gravitational contraction toward its longest axis. The sub-parsec segment we focused on is of great interest since it presents a variety of early-type objects of star formation that very likely
originated from the same parent filament. We showed a rare
example of changes of dust properties correlated with the evolutionary
stage of star formation. 

By means of the ratio map of the two NIKA bands we showed that, in the
Rayleigh-Jeans limit, the dust emissivity spectral index progressively decreases
from prestellar cores ($\beta_{\rm RJ}$$\sim$2) to
protostellar cores ($\beta_{\rm RJ}$$\sim$1) to a Class-II T-Tauri star ($\beta_{\rm
  RJ}$$\sim$0). Using temperature
radial profiles deduced from the {\it
  Herschel} data, $T_{\rm d}$, we also presented radial
profiles of the dust emissivity index, $\beta$, for two protostellar cores and one prestellar core, which were corrected for possible line-of-sight temperature
gradients through the Abel transform inversion technique. 

Our results showed that while a constant $\beta$ profile of about $2.4
\pm 0.3$ characterizes the prestellar core, decreasing $\beta$ profiles
toward the center are found for the two protostellar cores,
reaching central values between $\sim$$1$ and $\sim$$1.5$. A clear anticorrelation between $\beta$
and $T_{\rm d}$ is also observed, which is consistent with previous
observational works despite the lack of a theoretical framework able to accurately
model it. We concluded that the observed changes in $\beta$
with radius trace dust evolution from the prestellar to the protostellar stage, likely due to both grain-growth and
dust temperature effects in the mm range. 

In order to confirm this conclusion similar data should be
acquired. Now the advent of the NIKA2 instrument on the
IRAM-30m telescope \citep{Calvo2016} will give us the chance to enlarge the observed sample 
of objects at early stages of star formation and characterize their physical properties
at long wavelengths. Thanks to its polarization-sensitive capability at 1.2 mm \citep[see][]{Ritacco2017}, 
the NIKA2 camera will also allow us to
gain insight into the magnetic field structure of star-forming regions, especially at the scales 
where filaments fragment into prestellar cores, 
providing the community with an exceptional opportunity to enlighten
one of the most important but yet poorly explored aspects of the star formation process. 

\begin{acknowledgements} 
We would like to thank the IRAM staff for their support during the campaigns. 
The NIKA dilution cryostat has been designed and built at the Institut N\'eel. 
In particular, we acknowledge the crucial contribution of the Cryogenics Group, and 
in particular Gregory Garde, Henri Rodenas, Jean Paul Leggeri, Philippe Camus. 
This work has been partially funded by the Foundation Nanoscience Grenoble, the LabEx FOCUS ANR-11-LABX-0013 and 
the ANR under the contracts 'MKIDS', 'NIKA', and 'NIKA2SKY' (ANR-15-CE31-0017). 
This work has benefited from the support of the European Research Council Advanced Grant ORISTARS 
under the European Union's Seventh Framework Programme (Grant Agreement no. 291294).
We acknowledge fundings from the ENIGMASS French LabEx (R. A. and F. R.), 
the CNES post-doctoral fellowship program (R. A.),  the CNES doctoral fellowship program (A. R.) and 
the FOCUS French LabEx doctoral fellowship program (A. R.).
This research has made also use of data from the
{\it Herschel} Gould Belt survey (HGBS) project
(http://gouldbelt-herschel.cea.fr). The HGBS is a {\it Herschel} Key
Programme jointly carried out by SPIRE Specialist Astronomy Group 3
(SAG 3), scientists of several institutes in the PACS Consortium (CEA
Saclay, INAF-IFSI Rome and INAF-Arcetri, KU Leuven, MPIA Heidelberg),
and scientists of the Herschel Science Center (HSC).
P.~P. acknowledges support from the Funda\c{c}\~ao para a Ci\^encia e  
a Tecnologia of Portugal (FCT) through national funds  
(UID/FIS/04434/2013) and by FEDER through COMPETE2020  
(POCI-01-0145-FEDER-007672) and also by the fellowship  
SFRH/BPD/110176/2015 funded by FCT (Portugal) and POPH/FSE (EC).

\end{acknowledgements}

\bibliographystyle{aa}

\appendix

\section{Time coverage of the NIKA maps}\label{sec:app}
\begin{figure}
	\includegraphics[width=9.cm]{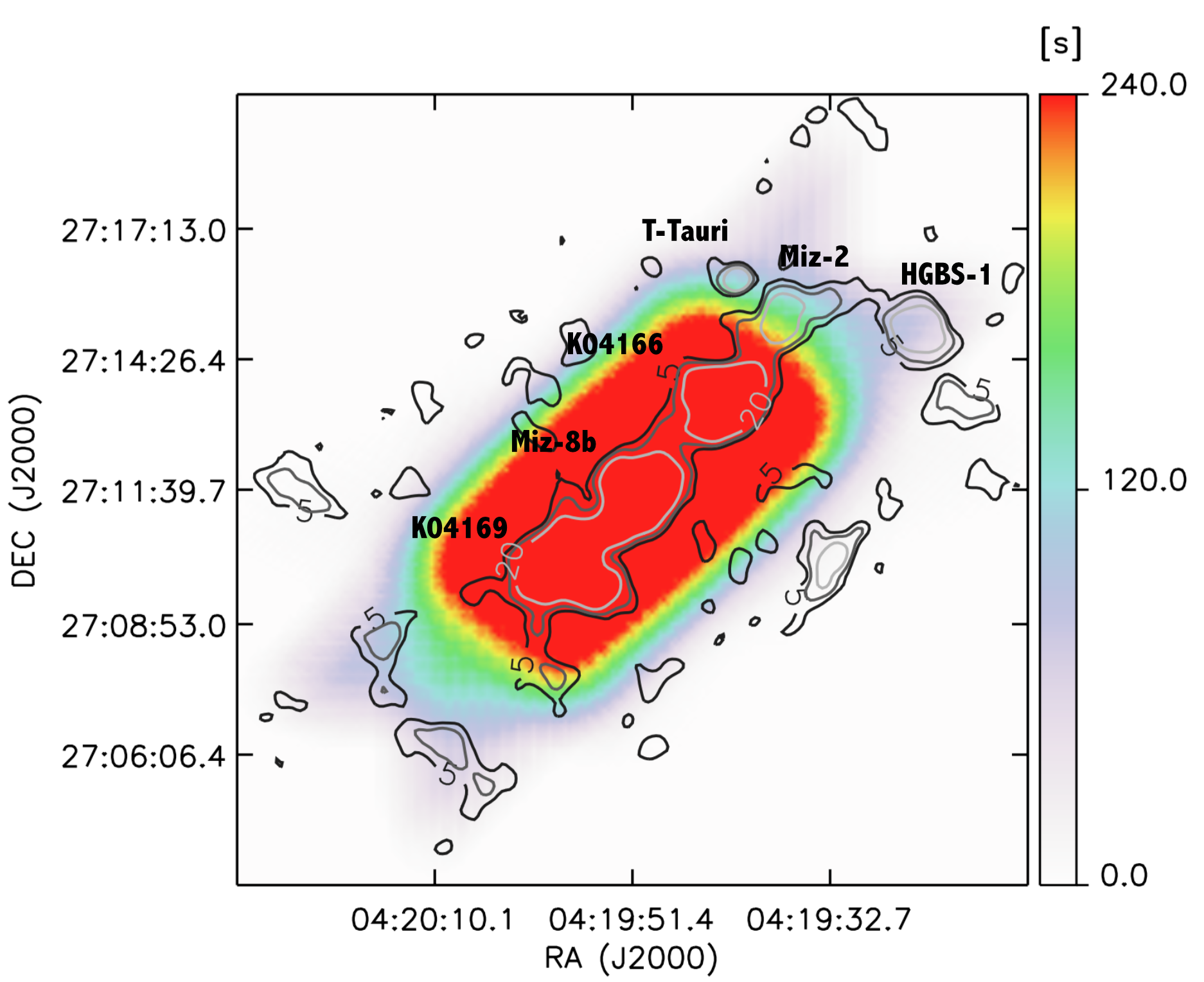} 
	\caption[]{Time-per-pixel map with contours of the 1.15 mm brightness
		as in the left panel of Fig.~\ref{fig:Nika}.}
	\label{fig:timemap}
\end{figure}

In this Appendix we show the time-per-pixel map that corresponds
to the NIKA observations presented in the paper. In
Fig.~\ref{fig:timemap} the time-per-pixel is shown in units of seconds
with the 1.15 mm brightness in contours as in the left panel of
Fig.~\ref{fig:Nika}. We use this map to justify our choice to detail
the analysis of dust properties only for the three sources that
were longer observed (i.e. K04166, K04169, and Miz-8b). These objects are
also those lying at the center of the field of view, where systematic
effects due to the scanning strategy are less significant.   


\end{document}